\documentclass[groupedaddress,
 reprint,
 amsmath,amssymb,
 aps,
 prd,
]{revtex4-1}

\usepackage{graphicx}
\usepackage{dcolumn}
\usepackage{bm}
\usepackage{hyperref}
\usepackage[utf8]{inputenc}
\usepackage[capitalise]{cleveref}
\usepackage{aas_macros}
\usepackage{color}

\begin{document}

\title{Baryon-driven growth of solitonic cores in fuzzy dark matter halos}

\author{Jan Veltmaat}
\author{Bodo Schwabe}
\author{Jens C. Niemeyer}

\affiliation{%
 Institut f\"ur Astrophysik\\
 Universit\"at G\"ottingen
}

\date{\today}

\begin{abstract}
We present zoom-in simulations of fuzzy dark matter (FDM) halos including baryons and star-formation with sufficient resolution to follow the formation and evolution of central solitons. We find that their properties are determined by the local dark matter velocity dispersion in the combined dark matter-baryon gravitational potential. This motivates a simple prescription to estimate the radial density profiles of FDM cores in the presence of baryons. As cores become more massive and compact if baryons are included, galactic rotation curve measurements are likely harder to reconcile with FDM.
\end{abstract}


\maketitle

\section{Introduction}

Ultralight scalar field models for dark matter \cite{Sin1994,Lopez2019}, also called fuzzy dark matter (FDM) \cite{Hu2000}, are phenomenologically interesting due to interference effects on length scales comparable to the de Broglie wavelength. If the scalar field mass is around $10^{-22}$ eV, one expects deviations from cold dark matter (CDM) on galactic scales while CDM predictions for cosmological large-scale structures are largely unaltered \cite{Hu2000, Marsh2015a, Hui2017}. Possible candidates for FDM are axion-like particles generically produced in string theories \cite{Arvanitaki2010,Marsh2015a}.

Detailed insight into the structure formed by FDM in the nonlinear regime can only be gained by numerical simulations. Cosmological dark-matter-only simulations solving the Schrödinger-Poisson (SP) equation show the formation of FDM halos with fluctuating interference patterns on scales of the de Broglie wavelength  \cite{Woo2009,Schive2014,Veltmaat2018}. Their cores consist of solitonic objects (Bose stars) whose attractive gravitational force is balanced by the repulsive effect of the scalar field gradient. Unlike the surrounding density peaks, it forms a coherent bound structure that remains stable apart from pulsating oscillations \cite{Veltmaat2018}.
Ref. \cite{Schive2014} empirically found a scaling relation between the mass of a FDM halo $M_\mathrm{vir}$ and the mass of its central soliton $M_\mathrm{sol}$ scaling as $M_\mathrm{sol} \sim M_\mathrm{vir}^{1/3}$. This relation heuristically follows from the equality of the spatial size of the soliton and the typical de Broglie wavelength of the host halo \cite{Schive2014a,Veltmaat2018} and can be explained by the saturation of mass growth by Bose-Einstein condensation \cite{Eggemeier2019}.

The presence of solitonic cores together with their core-halo mass relation defines the central dark matter distribution in FDM halos. Consequently, stellar kinematic data of dark matter dominated dwarf spheroidal and ultra-diffuse galaxies favour a specific value for the scalar field mass, which is typically around $m \sim 10^{-22}$ eV \cite{Schive2014,Marsh2015c,Chen2016,Calabrese2016,Gonzales-Morales:2016mkl,Broadhurst2019,Wasserman2019}. This value is in conflict with Lyman-$\alpha$ forest observations constraining the field mass to roughly $m > 10^{-21}$ eV \cite{Irsic2017,Kobayashi2017,Armengaud2017,Nori2019}. Less stringent constraints can be inferred from the high-$z$ galaxy luminosity function \cite{Bozek2015,Schive2016,Sarkar2016,Corasaniti2017,Lidz2018}. Galaxies with measured rotation curves yield even lower masses than inferred from dwarf spheroidals \cite{Bernal2018} or at least exclude the typical value $m \sim 10^{-22}$ eV  \cite{Bar2018,Robles2019}. In general, the scaling properties of solitons seem to disagree with   
observational data \cite{Safarzadeh2019,Deng2018}.

Most of the constraints on FDM cores so far rely on the core-halo relation derived from dark matter-only simulations. The effects of stars and supermassive black holes have been investigated analytically or in simplified numerical settings \cite{Chan2017,Bar2018,Bar2019,Yarnell2019,Emami2018}. In this paper, we present first results from simulations of galaxy formation with FDM together with standard prescriptions for hydrodynamics and star formation. Compared to similar simulations recently reported in Ref. \cite{Mocz2019}, we use a zoom-in technique to focus on smaller volumes and higher spatial resolution in order to explore the properties of solitonic cores in the presence of baryons for the first time. We specifically address the following questions: does a core still form in the presence of baryons and if it does, what determines its properties? 


\section{Methods}

\begin{figure}
\includegraphics[width=\columnwidth]{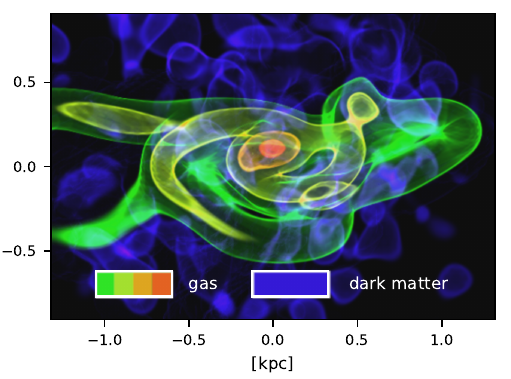}
\caption{Density volume rendering of the central region of halo 1 at $z=4.4$.} 
\label{fig:volumerendering}
\end{figure}

\begin{table*}
\begin{tabular}{c|c|c|c|c|c|c|c}
 halo & $z_\mathrm{f}$& $M_\mathrm{vir}$ [$10^{10}$ M$_\odot$]& $M_\mathrm{sol}^\mathrm{bar}/M_\mathrm{sol}^\mathrm{dm}$&$ M_\mathrm{tot}/M_\mathrm{dm}$&$v_\mathrm{sol}$ [km/s]& $v_\mathrm{c}$ [km/s]& $v_\mathrm{o}$ [km/s]\\\hline
1 & 4.0 & $1.05$ & 2.08 & 3.81 & 136 &138& 70\\
2 & 4.4 & $1.25$ & 3.00 & 3.07 & 159& 147 & 77
\end{tabular}
\caption{\label{tab:table1} Final values for redshift, virial mass, ratio of the soliton mass with baryons to the soliton mass without baryons given by the smoothed curve in \cref{fig:plot3}, ratio of total mass to dark matter mass within two times the half-density radius of the dark matter profile ($2\, r_{1/2}$) and velocities shown in \cref{fig:plot2} (soliton, central, outer)
for both halos. All values are taken from the FDM runs with baryonic physics included if not stated otherwise. 
} 
\end{table*}

We use the modified version of the public adaptive-mesh refinement (AMR) code \textsc{Enzo} \cite{Bryan2014} described in \cite{Veltmaat2018} for dark matter-only simulations. It employs a hybrid approach to solve the SP equation governing the dynamics of FDM,
\begin{align}
    i \hbar \left(\frac{\partial \Psi}{\partial t} + \frac{3}{2}H\Psi\right)&= - \frac{\hbar^2}{2m}\nabla^2 \Psi + V m \Psi \nonumber \cr
    \nabla^2 V &= 4 \pi G (\rho_\mathrm{tot} - \bar{\rho}_\mathrm{tot}) \,,
\end{align}
using a finite difference scheme for $\Psi$ on the finest AMR level while approximating the dynamics as a collisionless fluid on coarser levels with N-body particles.

\begin{figure}
\includegraphics[width=\columnwidth]{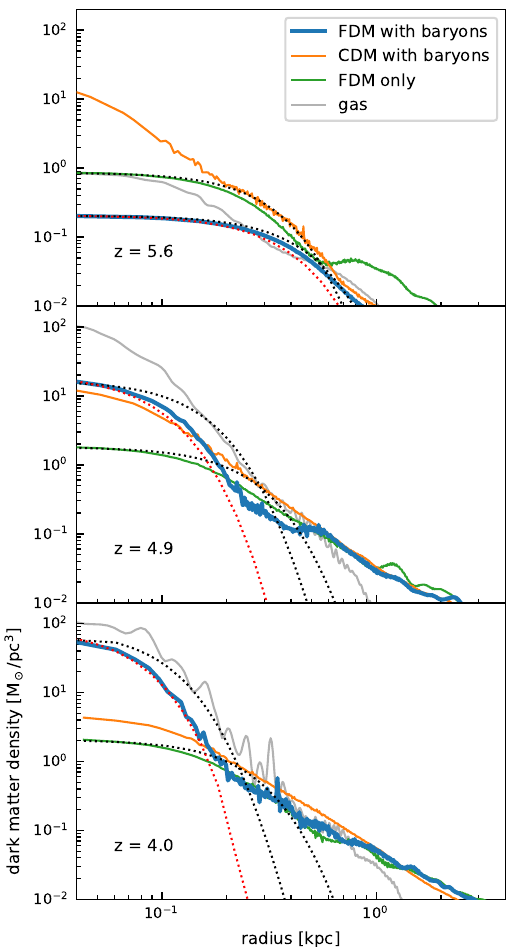}
\caption{Radial dark matter density profiles of halo 1 in all three runs at three different redshifts. The inner profile of the FDM run with baryons matches the modified FDM ground state solution (red dotted line) instead of the dark matter-only ground state solution (black dotted line). Also shown is the gas density profile of the FDM run with baryons.}
\label{fig:plot1}
\end{figure}

Here, we extend the simulations reported in \cite{Veltmaat2018} by additionally including baryonic physics as implemented into \textsc{Enzo} and described below. The total density $\rho_\mathrm{tot}$ with global average $\bar{\rho}_\mathrm{tot}$ entering on the right hand-side of the Poisson equation is then given by
\begin{align}
 \rho_\mathrm{tot}  = |\Psi|^2 + \rho_\mathrm{gas}+\rho_\mathrm{stars} \nonumber
\end{align}
with individual contributions from FDM, gas and stars respectively.
The hydrodynamic equations are integrated by the \textsc{Zeus} solver \cite{Stone1992} on the same AMR grid hierarchy as dark matter. We include non-equilibrium cooling solving the rate equations for H, H$^+$, He, He$^+$, He$^{++}$ and equilibrium cooling for metals using a lookup table from \cite{Glover2007} assuming solar abundances. We adopt a uniform metagalactic UV background computed in \cite{Haardt2012} which is gradually ramped up from zero at $z = 7.00$ to full strength at $z = 6.75$. For modelling star formation, we use the algorithm from \cite{Cen1992} adapted to \textsc{Enzo} as described in \cite{Bryan2014}. 
Stars form according to the standard criteria and delayed cooling is used to prevent artificial overcooling (see \cref{sec:sf}).

Our simulations use the same numerical and cosmological parameters as in \cite{Veltmaat2018} apart from changes related to the inclusion of baryons, i.e. 
$h= 0.7$, $\Omega_\Lambda = 0.75$, $\Omega_m = 0.25$, $\Omega_b = 0.05$, and the scalar field mass $m = 2.5 \times 10^{-22}$ eV. They start at a redshift of $z = 60$ from initial conditions generated by \textsc{Music} \cite{Hahn2011} with dark matter and baryon transfer functions computed by Axion\textsc{Camb} \cite{Hlozek2017}. With a total physical box size of 2.5 Mpc/$h$ covered by a root grid of $512^3$ cells and five additional refinement levels (two of them initial and static and three following the selected halo), we reach a comoving resolution of $\Delta x = 218$ pc in the region of the selected halo. 

The following analysis is based on multiple simulations of two halos, chosen in low-resolution CDM simulations and re-simulated with grids of higher resolution centered on the halo. For each halo, three high-resolution runs are conducted: (1) collisionless N-body dynamics appropriate for CDM including  baryonic physics, (2) FDM without baryons as in \cite{Veltmaat2018}, and (3) a full FDM simulation including baryonic physics. Ionizing background radiation prevents gas accretion in halos below $M_\mathrm{vir} \approx 10^9$ M$_\odot$ \cite{Efstathiou1992,Okamoto2008}, hence more massive halos with virial masses of $M_\mathrm{vir} \simeq 10^{10}$ M$_\odot$ were selected. For the purposes of studying the gravitational impact of baryonic physics on the formation of the solitonic FDM core, the initial phase of gas cooling and supernova feedback until the gas has reached quasi-hydrostatic equilibrium is most relevant. The simulations were therefore terminated at a redshift of $z \simeq 4$, allowing proper spatial resolution of the soliton until the end.

The parameters characterizing both halos are summarized in \cref{tab:table1}.


\section{Results}

The selected halos virialize at $z \approx 6$ and gradually grow without mergers to their final virial mass of $M_\mathrm{vir} \simeq 10^{10}$ M$_\odot$ at $z_\mathrm{f} \simeq 4$. Below we focus on the first halo, shown in \cref{fig:volumerendering}, as the second one yields comparable results.

At $z_\mathrm{f}$, the total stellar mass of halo 1 is $M_* = 1.5 \times 10^6$ M$_\odot$ and $M_* = 7.1 \times 10^6$ M$_\odot$ in the FDM and CDM runs, respectively.
The difference of more than a factor of four is caused by different numerical timesteps
affecting the star formation rate via the star-formation subgrid-scale model \cref{eq:starformation}. Given the statistical nature of the problem, a quantitative comparison of star formation histories in CDM and FDM scenarios would require a large number of halos followed across the peak of their star-forming activity. Studying the formation of solitonic cores in the presence of baryons, which is the focus of this work, mainly requires feedback in order to avoid runaway cooling and doesn't depend on the exact amount of star formation. 

\cref{fig:plot1} compares radial density profiles of halo 1 at different redshifts. The gas density profiles of the CDM and FDM runs  do not differ significantly in the central region for $z \lesssim 5.4$. The earliest redshift $z=5.6$ corresponds roughly to the time when the gaseous component has cooled and collapsed allowing star formation in the center.
At this time, the CDM run exhibits the highest central dark matter density with a cuspy profile, while the FDM runs show cores with relatively large radii in agreement with the core-halo mass relation, $M_\mathrm{sol} \sim M_\mathrm{vir}^{-1/3}$  \cite{Schive2014}. The FDM halo with baryons has a lower overall density than the FDM only halo because the baryon pressure delays its collapse.

\begin{figure}
\includegraphics[width=\columnwidth]{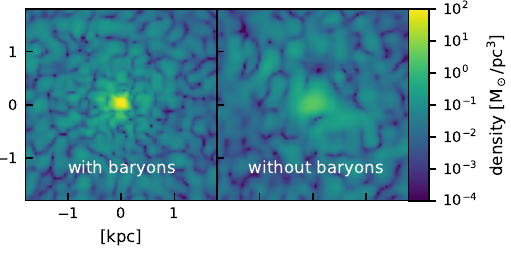}
\caption{Slices of dark matter density with and without baryons at $z=4.0$.} 
\label{fig:slices}
\end{figure}
At $z=4.0$, the FDM central density is more than one order of magnitude higher with baryons than without, exceeding the central dark matter density in the CDM run. Comparing slices of the FDM density fields with and without baryons, \cref{fig:slices} confirms the more compact core in the former.
It is surrounded by granular structures with similar spatial extent. This result agrees with Ref.\ \cite{Chan2017} who found that the addition of stars to an FDM halo leads to a more prominent core. As in \cite{Chan2017}, we observe a deviation of the core density profile from the soliton solution in the vacuum.

The density profile of the dark matter core with baryons is well described by a ground state solution of the SP equation taking into account the additional gravitational potential sourced by the baryon density. Similar solitonic solutions with an additional gravitating component have been investigated in the context of supermassive black holes \cite{Bar2018,Yarnell2019} and galactic disks \cite{Bar2019}. Our simulations confirm and extend these results by providing evidence that modified ground state solutions are approached dynamically in the presence of baryonic feedback (\cref{sec:sol_baryons}).

Ground state solutions of the SP equation are uniquely determined by their total mass or, alternatively, their central density $\rho_\mathrm{c}$. Dark-matter-only simulations show a relation between the masses of FDM halos and their solitonic cores which follows from an equilibrium of their virial temperatures \cite{Schive2014,Veltmaat2018,Bar2018,Eggemeier2019}. 

\begin{figure}
\includegraphics[width=\columnwidth]{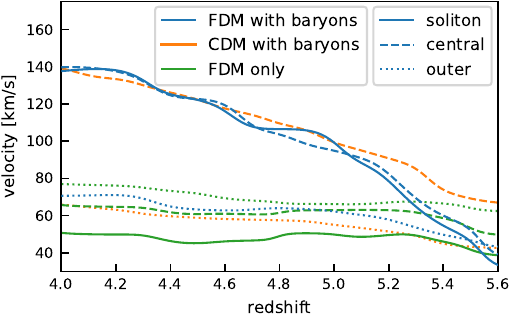}
\caption{Redshift evolution of velocity dispersion in the soliton (only in the runs with FDM), its local environment, and further outside at $x_\mathrm{vir}/2$ in the three different runs (see \cref{sec:vel_disp}). Data points are smoothed with a Gaussian filter with standard deviation of $\sigma_z = 0.08$ in redshift space.} 
\label{fig:plot2}
\end{figure}
In order to test whether this temperature correspondence holds in the presence of a baryonic component, we compare the virial velocity of the simulated cores with the local dark matter velocity dispersion in their environment in  \cref{fig:plot2} (see also \cref{sec:vel_disp}). 
Our simulations indeed verify that the baryonic gravitational potential gives rise to a radially stratified velocity distribution and 
confirm that the velocity dispersion of the soliton (and thus its virial temperature) closely follows the temperature of its immediate surrounding. In all runs with baryons, the central velocity dispersion differs by up to a factor of two from the velocity dispersion at $x_\mathrm{vir}/2$, whereas in the dark matter only runs, the velocity dispersion in the center and at the outer radius are similar. The similarity of the CDM and the FDM runs, both with baryons, suggests that the radial velocity distribution is a generic result of the accumulation of gas in the center unrelated to the distinctive features of FDM.

\begin{figure}
\includegraphics[width=\columnwidth]{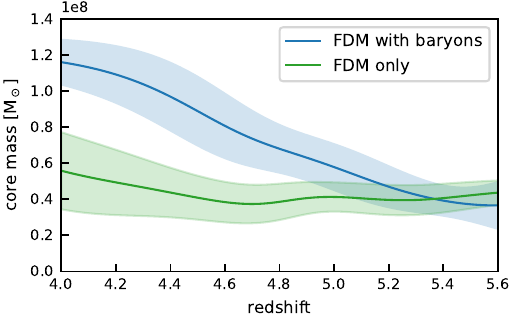}
\caption{Evolution of the core mass, defined as the dark matter mass within $\lambda_\mathrm{dB}/4$ from the center, where $\lambda_\mathrm{dB}$ is the \mbox{de Broglie} wavelength corresponding to $v_\mathrm{sol}$. The lines show the Gaussian filtered data points with $\sigma_z=0.2$. The shaded regions represent the corresponding standard deviations.} 
\label{fig:plot3}
\end{figure}
The balance of core and halo velocity dispersions is accompanied by a growth of the core mass, as illustrated in \cref{fig:plot3}. In the presence of baryons, the cores grow by more than a factor of two. In contrast, there is no clear sign of mass growth if baryons are absent, confirming previous results \cite{Veltmaat2018}.

\begin{figure}
\includegraphics[width=\columnwidth]{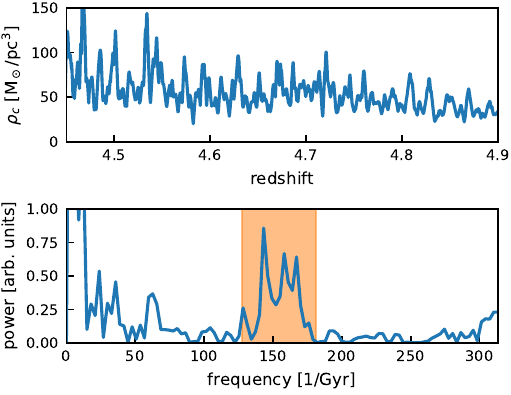}
\caption{Top: Evolution of the central density in the FDM run with baryons. Bottom: Frequency spectrum of the time series above. The orange-shaded region marks the expected quasinormal frequency under the influence of baryons. Its boundaries are computed using the minimum and maximum central density averaged over several oscillation periods.} 
\label{fig:plot4}
\end{figure}

Previous dark matter-only simulations \cite{Veltmaat2018} found that central solitonic cores are in excited states oscillating with their quasi-normal frequency \cite{Guzman2004}
\begin{align}
    f = 10.94\left(\frac{\rho_{c}}{10^{9}M_{\odot}\text{kpc}^{-3}}\right)^{1/2}\text{ Gyr}^{-1} \nonumber
\end{align}
inversely proportional to the free-fall time of the inner halo region with central density $\rho_\mathrm{c}$. If a baryonic component is present, the free-fall time depends on the total density $\rho_\mathrm{tot}$. Thus, assuming that the proportionality between quasi-normal period and free-fall time holds, one expects the frequency $f$ to increase by a factor $\approx \sqrt{\rho_\mathrm{tot}/\rho_\mathrm{c}}$. As shown in \cref{fig:plot4}, we indeed find that the ground state configurations in our FDM simulations with baryons oscillate with the frequency $f$ multiplied by the square root of total mass over dark matter mass within $2\, r_{1/2}$ averaged over time. 

\section{Conclusions}

Using cosmological hydrodynamical simulations including baryonic feedback, we find that the formation of central solitonic cores remains a robust prediction for FDM halos.
However, the core-halo mass relation found in dark matter only simulations \cite{Schive2014} is altered by two effects. Firstly, the accumulation of gas and stars leads to an increased dark matter velocity dispersion in the center. This effect is also found in CDM simulations and not special to FDM. As in the case of pure dark matter simulations, the velocity dispersion of the solitonic FDM core follows the ambient dark matter velocity dispersion, but due to the radially varying velocity dispersion profile the ambient dark matter velocity is now different from the virial velocity of the halo. Secondly, taking into account the gravitational effect of the baryons gives rise to a modified ground-state solution of the SP equation. This modified profile has a different mass-radius relation than a pure FDM soliton without baryons. Note that the two effects are opposite in the sense that increasing the velocity of the core with a fixed baryon profile increases the core mass, whereas increasing the baryon density with a fixed core velocity decreases it.

According to this result, core profiles in galaxies with non-negligible central amounts of baryons can be predicted in the following way: given the baryonic contribution to the gravitational potential, one solves for the ground state solution with velocity dispersion matching the velocity dispersion of the inner halo region. Obtaining the central dark matter velocity dispersion, however, requires numerical simulations or Jeans modelling. Ref.s \cite{Bar2018,Bar2019} used nearly the same approach, but they fixed the core velocity by the virial velocity of the halo, thus not taking into account the first of the two effects explained above.

The magnitude of change of the core-halo mass relation depends crucially on the central baryonic density profile. One therefore expects centrally baryon-dominated galaxies, especially those containing stellar bulges or supermassive black holes, to be more strongly affected than dwarf spheroidal galaxies. In our simulations, the central baryon profile is subject to uncertainties resulting from the sub-grid modelling of stellar feedback. Taking the simulated baryon distributions at face value, our more massive and more compact cores make FDM even harder to reconcile with galactic rotation curve measurements, because \cite{Bar2018,Robles2019} found that cores predicted by the standard core-halo mass relation with FDM masses around $10^{-22}$ eV are already too dense. It may be fruitful to revisit these constraints using the prescription above together with baryon profiles inferred from observations.

 
\section*{Acknowledgements}
We thank Benedikt Eggemeier, Erik Lentz, Doddy Marsh, and Philip Mocz for useful discussions. We acknowledge support by the Deutsche Forschungsgemeinschaft. JCN acknowledges funding by a Julius von Haast Fellowship Award provided by the New Zealand Ministry of Business, Innovation and Employment and administered by the Royal Society of New Zealand. Computations described
in this work were performed using the publicly-available
\textsc{Enzo} code (http://enzo-project.org) with resources provided by the North-German Supercomputing Alliance (HLRN). We acknowledge the \textsc{yt} toolkit \cite{Turk2011} that was used for our analysis of
numerical data.


\appendix

\section{Modeling star formation and feedback}
\label{sec:sf}

If in any timestep $\Delta t$ a grid cell with side length $\Delta x$ meets the following criteria, it will immediately form a star particle: (1) overdensity is greater than 100, (2) convergent gas flow, (3) cooling time is less than the dynamical time, (4) the cell is Jeans unstable, (5) the resulting star particle would have a mass larger than 5000 M$_\odot$. Mass is transferred from the gas density to the newly formed star particle according to
\begin{align}
\label{eq:starformation}
     m_* = f_* \frac{\Delta t}{t_\mathrm{dyn}} \rho_\mathrm{gas} (\Delta x)^3\,,
\end{align}
where $t_\mathrm{dyn} = \sqrt{3 \pi/(32 G \rho_{tot})}$ is the free-fall time.
We adopt a star formation efficiency of $f_* = 0.01$. A star particle returns $\epsilon_\mathrm{SN} = 10^{-5}$ of its rest mass energy in the form of thermal energy to the gas cell it resides in. The process is distributed over the timescale $t_\mathrm{dyn}$ evaluated at its formation. 

This thermal-only prescription of stellar feedback is known to suffer from overcooling if the spatial resolution is insufficient. Feedback energy is then radiated away too efficiently, leading to extreme collapse of gas and, in turn, unrealistically high star formation rates \cite{Katz1992}. To prevent this, we artificially turn off cooling in cells that receive feedback energy for 50 Myr after the star particle was formed. This ad-hoc method was shown to achieve desired results in the sense that catastrophic gas collapse is prevented and star formation rates drop to realistic values (see \cite{Hummels2012} and references therein). 

\section{Solitonic solutions with baryonic gravitational potential}
\label{sec:sol_baryons}

We assume spherical symmetry and approximate the baryonic density profile by
\begin{align}
    \rho_b(r) = \frac{\rho_{b0}}{1 + (r/r_{b0})^3}
\end{align}
with the two parameters $\rho_{b0}$ and $r_{b0}$ chosen to match the density profile found in the simulations. The ground state solution is a function $\Psi(\mathbf{x},t) = e^{-i \gamma t/ \hbar} \phi(|\mathbf{x}|)$ with the lowest energy $\gamma$ for a given total mass, where $\phi$ solves the following set of ordinary differential equations \cite{Bar2018}:
\begin{align}
    \label{eq:ground_sol}
   \frac{1}{r} \frac{\partial^2}{\partial r^2}(r\phi (r)) &= 2 (\frac{m^2}{\hbar^2}V(r) - \frac{m}{\hbar^2}\gamma) \phi(r) \cr
   \frac{1}{r} \frac{\partial^2}{\partial r^2}(r V(r)) &= 4\pi G (|\phi|^2(r)+\rho_b(r))
\end{align}
Solving \cref{eq:ground_sol} numerically with the boundary conditions $\phi(0) = \sqrt{\rho_c}$ and $\frac{\partial}{\partial r} \phi (0) = 0$, we find a unique value for $\gamma$ such that $\phi(\infty) = 0$ without zero-crossings (see \cite{Guzman2004}).

\section{Velocity dispersion of the core and its local environment}
\label{sec:vel_disp}

The FDM velocity dispersion is given by
\begin{align}
\label{eq:vrms}
    \sigma_v^2 = 2 \frac{K}{M}\, ,
\end{align}
where
\begin{align*}
\label{eq:kinenergy}
    M &= \int \text{d}^3\mathbf{x} |\Psi|^2 \qquad \mbox{and} \cr
    K &= \frac{\hbar^2}{2m^2} \int \text{d}^3\mathbf{x} |\nabla \Psi|^2
\end{align*}
are the total mass and kinetic energy of the field $\Psi$.

To determine the velocity dispersion of the cores, we integrate $M$ and $K$
to a radius of $2\, r_{1/2}$, where $r_{1/2}$ is defined as the radius where the angular-averaged density $\bar{\rho}(r)$ has dropped to $\rho_c/2$. The result for \cref{eq:vrms} is not very sensitive to this choice. 

Local velocity dispersions in the environment are determined by evaluating  \cref{eq:vrms} for the field
\begin{align}
    \tilde{\Psi}(\mathbf{x}) = \exp\left[-\frac{ (|\mathbf{x}|-f x_\mathrm{vir})^2}{2 \Sigma^2}\right] \frac{\Psi(\mathbf{x})}{\sqrt{\bar{\rho}(|\mathbf{x}|)}}  \,.
\end{align}
Dividing $\Psi$ by its angular mean amplitude erases the core feature but leaves the granular structure in the rest of the halo intact. The exponential serves as a window function at different radii, which are expressed here as fractions $f$ of the virial radius $x_\mathrm{vir}$.

We also compute the velocity dispersion of the N-body particles in the CDM runs. To this end, we multiply the particle masses with the same window function as above:
\begin{align}
    \sigma_v^2 = \frac{\Sigma_i \exp\left[-(|\mathbf{x_i}|-f x_\mathrm{vir})^2/(2\Sigma^2)\right] m_i v_i^2}{\Sigma_i \exp\left[- (|\mathbf{x_i}|-f x_\mathrm{vir})^2/(2\Sigma^2) \right] m_i} \,.
\end{align}

\Cref{fig:plot2} uses $f = 0, 0.5$ for the inner and outer radius, respectively. We choose $\Sigma = 0.4$ kpc in order to include several de Broglie wavelengths while staying inside of the virial radius, $x_\mathrm{vir}\sim 25$ kpc.

\end{document}